\UseRawInputEncoding
\documentclass[journal,twoside,web]{ieeecolor}
\usepackage{tmi}
\usepackage{cite}
\usepackage{amsmath,amssymb,amsfonts}
\usepackage{algorithmic}
\usepackage{graphicx}
\usepackage{textcomp}

\usepackage{hyperref}
\usepackage{booktabs}
\usepackage{multirow}

\usepackage{etoolbox}
\makeatletter
\@ifundefined{color@begingroup}%
{\let\color@begingroup\relax
	\let\color@endgroup\relax}{}%
\def\fix@ieeecolor@hbox#1{%
\hbox{\color@begingroup#1\color@endgroup}}
\patchcmd\@makecaption{\hbox}{\fix@ieeecolor@hbox}{}{\FAILED}
\patchcmd\@makecaption{\hbox}{\fix@ieeecolor@hbox}{}{\FAILED}

\def\BibTeX{{\rm B\kern-.05em{\sc i\kern-.025em b}\kern-.08em
    T\kern-.1667em\lower.7ex\hbox{E}\kern-.125emX}}
\begin{document}

	\title{FUGC: Benchmarking Semi-Supervised Learning Methods for Cervical Segmentation}
	\author{Jieyun Bai, Yitong Tang, Zihao Zhou, Mahdi Islam, Musarrat Tabassum, Enrique Almar-Munoz, Hongyu Liu, Hui Meng, Nianjiang Lv, Bo Deng, Yu Chen, Zilun Peng, Yusong Xiao, Li Xiao, Nam-Khanh Tran, Dac-Phu Phan-Le, Hai-Dang Nguyen, Xiao Liu, Jiale Hu, Mingxu Huang, Jitao Liang, Chaolu Feng, Xuezhi Zhang, Lyuyang Tong, Bo Du, Ha-Hieu Pham, Thanh-Huy Nguyen, Min Xu, Juntao Jiang, Jiangning Zhang, Yong Liu, Md. Kamrul Hasan, Jie Gan, Zhuonan Liang, Weidong Cai, Yuxin Huang, Gongning Luo, Mohammad Yaqub and Karim Lekadir
    \thanks{Jieyun Bai, Yitong Tang, Zihao Zhou, Bo Deng, Yu Chen, and Zilun Peng are with Department of Cardiovascular Surgery, The First Affiliated Hospital, Jinan University, China (jbai996@aucklanduni.ac.nz; zzh001217@163.com; 1546421578@qq.com;
db0725@stu2024.jnu.edu.cn; biabuluo@stu2024.jnu.edu.cn; maxpeng@stu2024.jnu.edu.cn) Md. Kamrul Hasan is with Imperial College London, UK (k.hasan22@imperial.ac.uk). Jie Gan, Zhuonan Liang, and Weidong Cai are with University of Sydney, Australia (jie.gan.ai@gmail.com; zlia3085@uni.sydney.edu.au; tom.cai@sydney.edu.au). Mohammad Yaqub is with Mohamed bin Zayed University of Artificial Intelligence, United Arab Emirates (mohammad.yaqub@mbzuai.ac.ae). Juntao Jiang, Jiangning Zhang, and Yong Liu are with Zhejiang University, China (juntaojiang@zju.edu.cn; 186368@zju.edu.cn; yongliu@iipc.zju.edu.cn). Thanh-Huy Nguyen and Min Xu are with Carnegie Mellon University, United States (bihuyz@gmail.com; mxu1@cs.cmu.edu). Mahdi Islam, Musarrat Tabassum, and Enrique Almar-Munoz are with Medical University of Innsbruck, Austria (mahdiiut079@gmail.com; tabassummusarrat212@gmail.com; enrique.almar@i-med.ac.at). Hongyu Liu, Hui Meng, and Nianjiang Lv are with University of Chinese Academy of Sciences, China (liuhongyu24@mails.ucas.ac.cn; huimeng@ucas.ac.cn; lvnianjiang23@mails.ucas.ac.cn). Yusong Xiao and Li Xiao are with University of Science and Technology of China, China (ysxiao@mail.ustc.edu.cn; xiaoli11@ustc.edu.cn). Nam-Khanh Tran, Dac-Phu Phan-Le, Hai-Dang Nguyen and Ha-Hieu Pham are with University of Science, Viet Nam National University Ho Chi Minh City, Vietnam (21120015@student.hcmus.edu.vn; 21120111@student.hcmus.edu.vn; nhdang@selab.hcmus.edu.vn; hieuphamha19@gmail.com). Xiao Liu and Jiale Hu are with Nanyang Institute of Technology, China (2215925735@nyist.edu.cn; 2215925709@nyist.edu.cn). Mingxu Huang, Jitao Liang, and Chaolu Feng are with Northeastern University, China (hmx1113953971@163.com; 2472127@stu.neu.edu.cn; fengchaolu@cse.neu.edu.cn). Xuezhi Zhang, Lyuyang Tong, and Bo Du are with Wuhan University, China (zhangxuezhi@whu.edu.cn; Lyuyangtong@whu.edu.cn; dubo@whu.edu.cn). Yuxin Huang is with Southern Medical University, China (15521280986@163.com). Gongning Luo is with Harbin Institute of Technology, China (luogongning@hit.edu.cn). Karim Lekadir is with Universitat de Barcelona, Spain (karim.lekadir@ub.edu). This work was supported by Guangzhou Municipal Science and Technology Bureau (No.2025B03J0127 and 2024B03J1283) and the ERC-funded AIMIX project (No.101044779). Corresponding author: Jieyun Bai and Yuxin Huang.}}
	\maketitle
	\begin{abstract}
Accurate segmentation of cervical structures in transvaginal ultrasound (TVS) is critical for assessing the risk of spontaneous preterm birth (PTB), yet the scarcity of labeled data limits the performance of supervised learning approaches. This paper introduces the Fetal Ultrasound Grand Challenge (FUGC), the first benchmark for semi-supervised learning in cervical segmentation, hosted at ISBI 2025. FUGC provides a dataset of 890 TVS images, including 500 training images, 90 validation images, and 300 test images. Methods were evaluated using the Dice Similarity Coefficient (DSC), Hausdorff Distance (HD), and runtime (RT), with a weighted combination of 0.4/0.4/0.2. The challenge attracted 10 teams with 82 participants submitting innovative solutions. The best-performing methods for each individual metric achieved 90.26\% mDSC, 38.88 mHD, and 32.85 ms RT, respectively. FUGC establishes a standardized benchmark for cervical segmentation, demonstrates the efficacy of semi-supervised methods with limited labeled data, and provides a foundation for AI-assisted clinical PTB risk assessment.
 	\end{abstract}
	
	\begin{IEEEkeywords}
Foundation Models, Fetal Ultrasound, Semi-supervised Learning, Self-supervised Learning, PTB, SAM, UniMatch, DINO, Cervical Length
	\end{IEEEkeywords}
	
	\section{Introduction}
	\label{sec:introduction}
%\subsection{Clinical Background}
\IEEEPARstart{T}{he} uterus is central to human reproduction, supporting fetal development and adapting dynamically across life stages, particularly during pregnancy\cite{xu2020uterus}. As a hormone-responsive fibromuscular organ, it must maintain quiescence until term and then transition to coordinated contractions for childbirth. Timely activation of both the uterus and cervix is essential: premature activation can cause preterm birth (PTB)\cite{2008Long}, a major contributor to neonatal mortality and lifelong morbidity, whereas delayed activation may lead to dystocia and related complications. PTB also imposes substantial social and economic burdens\cite{2008Long}. Effective risk-stratification tools, such as cervical length measurement, are therefore critical. The ISUOG Practice Guidelines underscore transvaginal sonographic (TVS) cervical length assessment as a key predictor of spontaneous PTB\cite{2022ISUOG}, highlighting the ongoing need for improved methods to enhance clinical outcomes.

Accurate cervical segmentation is essential for uterine functional assessment, providing precise measurements of cervical shape, length, and other contraction-related parameters critical for predicting spontaneous PTB\cite{2022ISUOG}. Clinically, these biomarkers guide risk stratification and timely intervention\cite{2022ISUOG}. Automated segmentation further enables objective, consistent longitudinal monitoring of cervical changes\cite{farras2024real}, improving decision-making while reducing clinical workload.

\subsection{Challenges}
With advances in supervised deep learning, numerous state-of-the-art ultrasound segmentation methods have been developed\cite{fiorentino2023review}. However, supervised models remain vulnerable to overfitting, particularly when labeled data are limited—an issue especially pronounced in medical ultrasound imaging, where data scarcity arises from the rarity of certain conditions, privacy constraints, and the high cost of expert annotation. Although strategies such as model simplification, regularization, and diversified augmentation\cite{fiorentino2023review} can mitigate these issues, their effectiveness still lags behind models trained on large, well-annotated datasets. Thus, developing approaches that maintain high performance with limited labels is increasingly critical for advancing clinical decision support.

Semi-supervised learning offers a promising direction by leveraging both scarce labeled data and abundant unlabeled data. Core semi-supervised learning strategies include pseudo-labeling and consistency regularization to stabilize predictions under perturbations\cite{chen2022semi}. Recent improvements, such as uncertainty-aware refinement and cross-supervision\cite{ran2024pseudo}, further enhance pseudo-label quality. Nonetheless, semi-supervised learning still faces key challenges, including fluctuating pseudo-label reliability, sensitivity to label noise, and overfitting within single-stage training pipelines\cite{yang2025unimatch}. Addressing these limitations requires dynamic, stage-aware optimization and multi-stage training frameworks to strengthen robustness, feature abstraction, and generalization.

\begin{figure}[!t]
\centering
\includegraphics[width=3.0 in]{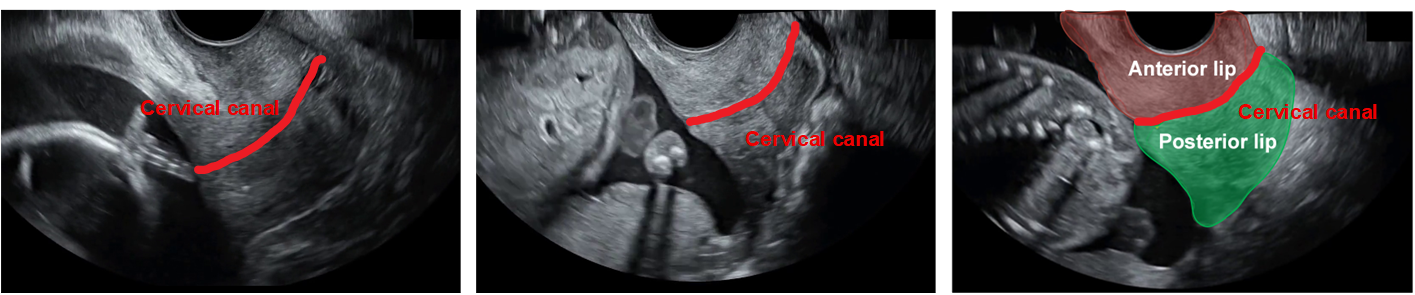}
\caption{The anterior and posterior lips are key anatomical structures for identifying the cervix, despite their wide variability in size and shape.}
\label{fig:fig1}
\end{figure}

Despite these advancements, a substantial gap persists in publicly available datasets and standardized evaluation frameworks for semi-supervised cervical segmentation on TVS images. Cervical segmentation itself presents unique challenges, including signal dropouts, artifacts, missing boundaries, distorted targets, attenuation, and shadows\cite{fiorentino2023review}, which hinder the performance of conventional segmentation approaches (\textcolor[RGB]{65, 105, 210}{\textbf{Fig.~\ref{fig:fig1}}}). These limitations highlight the need for more robust and reproducible semi-supervised learning methodologies.

\subsection{Motivation}
To advance research in semi-supervised cervical segmentation, the Fetal Ultrasound Grand Challenge (FUGC) was initiated to systematically evaluate and promote state-of-the-art semi-supervised learning methodologies, with the clinical goal of enabling more accurate cervical length measurement and improving early, precise screening for PTB risk\cite{scott2024virtual}. An open-access dataset was released, and the challenge was hosted on the Codabench platform (\textcolor{cyan}{\href{https://www.codabench.org/competitions/4781/}{https://www.codabench.org/competitions/4781/}}) during ISBI 2025, providing a standardized environment for benchmarking semi-supervised learning algorithms under consistent conditions. To ensure methodological rigor, the evaluation framework was developed in accordance with the Biomedical Image Analysis Challenges (BIAS) guidelines, promoting transparency, reproducibility, and high-quality reporting. Collectively, these efforts position FUGC as a catalyst for scalable, data-efficient cervical segmentation models, contributing to the quantitative study of uterine physiology and the development of AI tools for maternal–fetal medicine\cite{stock2023barriers}.

	\section{Related Work}
	\label{sec:introduction}
\subsection{Related Grand Challenges}
Over the past decade, public benchmark challenges have played a central role in advancing ultrasound image analysis by providing curated datasets and standardized evaluation protocols. Most challenges are hosted at major conferences such as MICCAI, ISBI, and PRECISE, and collectively span a wide range of tasks, including segmentation, detection, classification, registration, reconstruction, and biometry across diverse anatomical targets. Segmentation remains the dominant focus reflecting its fundamental importance for downstream ultrasound analysis workflows. Recent challenges have increasingly incorporated multi-task or multi-organ settings, highlighting a growing trend toward scalable, generalized ultrasound analysis frameworks\cite{bai2025psfhs,sappia2025acouslic}. Despite this progress, all existing ultrasound challenges rely exclusively on fully supervised learning, underscoring the absence of benchmarks that address label scarcity or evaluate semi-supervised methodologies\cite{jiang2024intrapartum,zhou2025ersr}, particularly for anatomically complex regions such as the cervix.

In recent years, semi-supervised learning challenges have been introduced across CT, X-ray, and MRI modalities, including FLARE, Semi-TeethSeg, and PI-CAI\cite{wang2026miccai}, collectively demonstrating the growing maturity of semi-supervised learning in medical imaging. Top-performing teams have adopted strategies such as iterative pseudo-labeling with uncertainty filtering, nnU-Net–based multi-stage pipelines, adaptive teacher–student frameworks, and hybrid detection–segmentation models guided by SAM-generated pseudo-labels, highlighting how pseudo-label refinement, consistency regularization, ensemble learning, and staged optimization enable strong performance with limited annotations. Despite these advances, no semi-supervised learning benchmark currently exists for ultrasound segmentation—an important gap given ultrasound’s real-time, low-cost clinical utility but also its susceptibility to operator dependence, artifacts, low contrast, and noise. These challenges are further amplified in transvaginal cervical imaging, where anatomical variability and restricted field-of-view hinder reliable boundary extraction. Establishing a semi-supervised benchmark for cervical ultrasound is therefore crucial for standardizing evaluation, fostering methodological innovation, and supporting PTB risk prediction.

\subsection{State-of-the-art Methods for Cervical Segmentation} 
Research on computational analysis of cervical ultrasound images has progressed steadily since early region-based methods in 2004 (\textcolor[RGB]{65, 105, 210}{\textbf{Table \ref{tab:related_research}}})\cite{wu2004novel}, followed by the introduction of deep learning architectures such as U-Net, FCN, Deeplabv3, residual U-Net, Transformer-based models, and multi-structure segmentation frameworks targeting the cervical canal, inner/outer boundaries, and adjacent organs\cite{wlodarczyk2019estimation,wlodarczyk2020preterm,dagle2022automated,pegios2023preterm,bones2024automatic,kwon2025deep,dagle2025generating}. However, unlike other ultrasound datasets, 2D cervical ultrasound datasets remain private, small-scale, and heterogeneous, limiting reproducibility and preventing systematic comparison across methods. Existing work also relies exclusively on fully supervised learning, which is impractical given the annotation burden and the significant imaging challenges unique to transvaginal cervical ultrasound, including acoustic shadowing, missing boundaries, anatomical variability, and limited field-of-view. The characteristics of cervical anatomy and the nature of TVS imaging make Dice Similarity Coefficient (DSC) and Hausdorff Distance (HD) appropriate for evaluating cervical segmentation performance. The absence of publicly accessible datasets and standardized benchmarks therefore represents a major bottleneck, preventing meaningful comparison across models and slowing methodological progress\cite{jiang2025pretraining}. 

\begin{table}[htbp]
    \centering
    \caption{Related Works of Cervical Segmentation. AP: Average Precision; JC: Jaccard similarity coefficient.}
    \renewcommand{\arraystretch}{1.2} 
    \resizebox{\columnwidth}{!}{%
        \begin{tabular}{p{0.2\textwidth}l p{0.2\textwidth}l p{0.2\textwidth}c p{0.2\textwidth}l p{0.2\textwidth}c}
            \toprule
            \textbf{Reference} & \textbf{Data Size} & \textbf{Data Dimensions} & \textbf{Methods} & \textbf{Results} \\
            \midrule
            Wu et al. (2004)\cite{wu2004novel} & 101 (Private) & 2D US & Region-based segmentation & - \\
            Włodarczyk et al.(2019)\cite{wlodarczyk2019estimation} & 359 (Private) & 2D US &  U-Net & JC=0.910 \\
            Wlodarczyk et al.(2020)\cite{wlodarczyk2020preterm} & 354 (Private) & 2D US &  U-Net & JC=0.923 \\
            Dagle et al.(2022)\cite{dagle2022automated} & 250 (Private) & 2D US &  Res-UNet & DSC=0.91 \\
            Pegios et al.(2023)\cite{pegios2023preterm} &  7862 (Private) & 2D US &  DTU-Net & - \\
            Boneš et al.(2024)\cite{bones2024automatic} &  298 (Public) & 3D US &  nnUNet & DSC=0.90  \\        
            Dagle et al. (2025)\cite{dagle2025generating}& 275 (Private) & 2D US &  Ensemble model & DSC=0.92 \\
            Kwon et al. (2025)\cite{kwon2025deep} & 474 (Private) & 2D US & CL-Net & DSC=0.95 \\
            Hwangbo et al.(2025)\cite{hwangbo2025occr} & 1625 (Private) & 2D US & O-CCR & AP=0.981 \\
            \bottomrule
        \end{tabular}%
    }
    \label{tab:related_research}
\end{table}

To address these gaps, we introduce the FUGC, the first semi-supervised benchmark for cervical ultrasound segmentation, detailing its dataset design, participating semi-supervised learning frameworks, and evaluation procedures following the BIAS methodology\cite{r25}. By establishing a standardized, reproducible, and globally accessible platform, FUGC aims to identify robust semi-supervised learning strategies for cervical segmentation and to catalyze broader advances in semi-supervised medical imaging.
  	
\section{Materials and Methods} 
\label{sec:methodology}
 \subsection{Dataset}
The dataset comprises 890 TVS images, including 500 training images (50 labeled and 450 unlabeled), 90 validation images, and 300 test images, acquired using GE Voluson E10 and GE Voluson E8 systems. Cohort demographics and clinical characteristics are summarized in \textcolor[RGB]{65, 105, 210}{\textbf{Table \ref{tab:cohort_summary}}}, with maternal age ranging from 18–43 years, a BMI of 25.15 ± 3.84, a prior cesarean section rate of 16.7\%, delivery occurring between 37+0 and 41+5 weeks, newborn weights ranging from 2040 to 4490 g, and a cesarean delivery rate of 34.6\%. Clinical data acquisition and verification were supported by Dr. Yuxin Huang. All scans were performed after bladder emptying with participants positioned in a low Fowler’s position, and operators were instructed to avoid post-processing artifacts while adjusting imaging parameters such as gain and frequency at their discretion. To protect patient privacy, all ultrasound images were anonymized by cropping out all patient-identifying text and metadata displayed on the original scans. After cropping, the remaining cervical region was stored at a uniform resolution of 544 × 336 pixels. Annotations of the anterior and posterior cervical lips were initially generated using a fine-tuned, ultrasound-adapted SAM model\cite{zhou2025segment} and subsequently refined using the PAIR software by an experienced obstetric ultrasonographer (Yuxin Huang, $>$10 years of expertise). The SAM model served only as an annotation aid and was not included as a participating method in the challenge; when evaluated independently, it achieved a DSC of 0.9831 and an HD of 5.3245, noting that it was trained on a dataset substantially larger and more diverse than the official challenge set. The data is available at: \textcolor{cyan}{\href{https://zenodo.org/uploads/16893174}{https://zenodo.org/uploads/16893174}} under the CC BY 4.0 license, and the study was approved by the Institutional Review Board (No. JNUKY-2022-019).

\begin{table}[htbp]
\centering
\caption{Summary of Cohort Demographics.}
   \renewcommand{\arraystretch}{1.2} 
    \resizebox{\columnwidth}{!}{%
\begin{tabular}{p{0.2\textwidth}l p{0.2\textwidth}l p{0.2\textwidth}l}
\toprule
\textbf{Category} & \textbf{Variables} & \textbf{ Description} \\
\midrule
\multirow{4}{*}{Maternal Demographics} 
& Maternal age (years) & 18--43 \\
& Weight (kg) & 48--90 \\
& Height (cm) & 152--172 \\
& Body mass index (BMI) & 25.15 $\pm$ 3.84 \\
& Previous cesarean section (\%) & 16.7\% \\
\midrule
\multirow{3}{*}{Pregnancy Outcomes}
& Delivery gestational age (weeks) & 37+0 to 41+5 \\
& Cesarean section (\%) & 34.6\% \\
& Newborn weight (g) & 2040--4490 \\
\bottomrule
\end{tabular}
}
\label{tab:cohort_summary}
\end{table}

\subsection{Evaluation Metrics and Ranking Scheme}
\subsubsection{Metrics} To quantify the segmentation performance, two well-established evaluation metrics are adopted: DSC and HD. To provide a deeper analysis of the segmentation performance across different regions, we extend these metrics into specific sub-categories. We use DSC\_P, DSC\_A, and DSC\_All, as well as HD\_P, HD\_A, and HD\_All to evaluate the segmentation performance separately for the anterior lip, posterior lip, and the overall cervix. To quantitatively evaluate the segmentation efficiency, we also measure the test-time segmentation runtime (RT) on the same hardware configuration (CPU: Xeon Silver 4210R, GPU: Quadro RTX 8000). The RT is calculated from the start time, which is defined as the loading of the first scan, to the stop time.
\subsubsection{Ranking Scheme}The ranking of this competition consists of two parts: the overall ranking and the individual rankings. Individual rankings are based on the rankings of each evaluation metric, while overall ranking is determined by the weighted scores across three metrics. 

RT was ranked in ascending order, with lower values indicating better performance, whereas DSC and HD 
metrics were analyzed through a multi-step ranking system. First, box-and-whisker plots were used to 
illustrate the distribution of each metric, including the minimum, maximum, median, and interquartile 
range, along with outliers. Second, blob plots were generated to assess ranking stability across metrics, 
where each blob’s area reflects the relative frequency of achieved ranks across 1{,}000 bootstrap samples; 
the median rank is marked by a black cross, with 95\% bootstrap intervals shown as black lines. Third, 
significance maps based on pairwise Wilcoxon signed-rank testing were constructed, where yellow cells 
indicate statistically superior performance of the column method over the row method (p $<$ 0.05), blue 
cells indicate no significant difference, and diagonal cells are intentionally left blank. Fourth, for each 
metric, five ranking strategies were computed: (1) mean-based aggregated scores (MeanTR), (2) median-based 
aggregated scores (MedianTR), (3) mean of per-metric ranks (RTMean), (4) median of per-metric ranks 
(RTMedian), and (5) significance-based rankings derived from Wilcoxon signed-rank test counts (TestBased). Importantly, the final ranking method adopted in the challenge was MeanTR, in which the final leaderboard positions were determined by averaging scores across all metrics 
and ranking teams according to these aggregated values. This decision was guided by a robustness analysis 
performed within ChallengeR (\textcolor{cyan}{\href{https://github.com/wiesenfa/challengeR}{https://github.com/wiesenfa/challengeR}}), which evaluates ranking stability through Kendall’s tau, a 
measure of rank correlation that quantifies the agreement between different ranking schemes. 

The ranking methodology for this competition was designed to prioritize segmentation accuracy over efficiency, reflecting the clinical importance of reliable cervical delineation. Accordingly, accuracy and efficiency were weighted at a 2:1 ratio, yielding final metric weights of 0.4, 0.4, and 0.2 for mDSC, mHD, and RT, respectively. The two complementary accuracy metrics—mDSC (averaging DSC\_A and DSC\_P) and mHD (averaging HD\_A and HD\_P)—were assigned equal importance, as they jointly capture region-overlap fidelity and boundary precision. To ensure comparability across metrics with different numerical ranges and optimization directions, all values were normalized using max–min scaling to the interval [0, 1], inverted where needed (for mHD and RT), and then mapped to a score range of 40–100 to encourage participation without affecting rankings. A 402 ms inference-time threshold, derived from the Baseline method (\textcolor{cyan}{\href{https://github.com/maskoffs/Fetal-Ultrasound-Grand-Challenge}{https://github.com/maskoffs/Fetal-Ultrasound-Grand-Challenge}}), was applied so that slower—but highly accurate—methods were not unduly penalized, with teams exceeding this threshold receiving the full RT score. To verify that the chosen weights did not bias rankings toward any particular metric or solution strategy, we conducted a comprehensive sensitivity analysis across the full weight space (see \textcolor{cyan}{Ranking Results} section). This analysis demonstrated that the selected weighting point (0.4, 0.4, 0.2) lies at the center of a stable, accuracy-dominant region.

\section{Challenge Entries}
\begin{figure}[!t]
		\centering
\includegraphics[width=3.5 in]{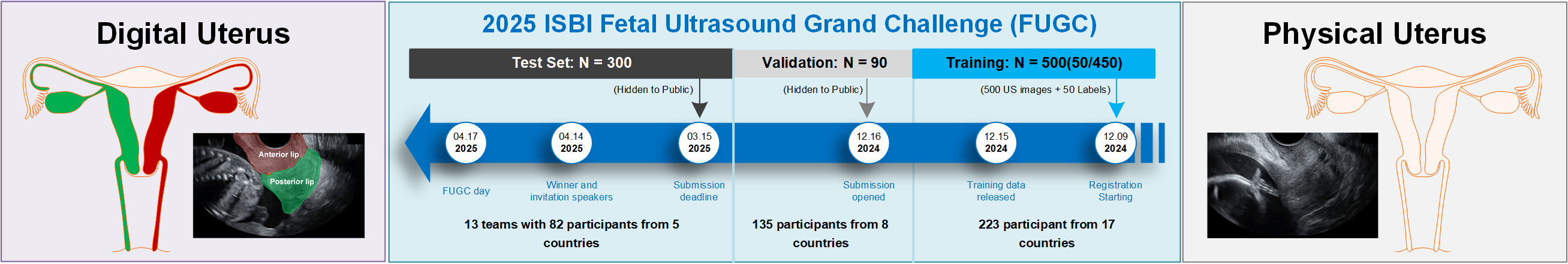}
\caption{The FUGC workflow, showing the segmentation focus, challenge timeline, participants, and dataset.}
\label{fig:fig2}
\end{figure}

In December 2024, a total of 223 participants from 17 countries registered for FUGC. Of these, 135 participants from 8 countries advanced to the validation phase. Ultimately, 13 teams comprising 82 participants from 5 countries completed the testing phase, with 10 teams receiving awards (\textcolor[RGB]{65,105,210}{\textbf{Fig.~\ref{fig:fig2}}}). The source code and corresponding methodological descriptions from the top 10 teams are summarized in \textcolor[RGB]{65,105,210}{\textbf{Table~\ref{tab:fugc2025_source_code}}}, and the complete collection of publicly released implementations has also been made available on GitHub at \textcolor{cyan}{\href{https://github.com/baijieyun/ISBI-2025-FUGC-Source-Code}{this repository}}.

\begin{table}[htbp]
    \centering
    \caption{Summary of source code and paper from the Top ten teams.}
    \renewcommand{\arraystretch}{1.2} % 调整行间距
    \resizebox{\columnwidth}{!}{%
        \begin{tabular}{c p{1.1\columnwidth}l}
            \toprule
            \textbf{Teams} & \textbf{Source code \& paper} & \textbf{References} \\
            \midrule
            T1 & https://zenodo.org/records/16013666 & Tran et al. 2025\cite{nam2025human} \\
            T2 & https://zenodo.org/records/16014145 & Pham et al. 2025\cite{ha2025fetal} \\
            T3 & https://zenodo.org/records/16014322 & Liu et al. 2025\cite{hongyu2025light} \\
            T4 & https://zenodo.org/records/16014436 & Zhang et al. 2025\cite{chensheng2025semi} \\
            T5 & https://zenodo.org/records/16015993 & Huang et al. 2025\cite{mingxu2025leveraging} \\
            T6 & https://zenodo.org/records/16016516 & Chen et al. 2025\cite{yu2025comt} \\
            T7 & https://zenodo.org/records/16016668 & Liu et al. 2025\cite{xiao2025vision} \\
            T8 & https://zenodo.org/records/16017032 & Xiao et al. 2025\cite{yusong2025hierarchical} \\
            T9 & https://zenodo.org/records/16017188 & Islam et al. 2025\cite{m2025semi} \\
            T10 & https://zenodo.org/records/16017509 & Jiang et al. 2025\cite{juntao2025semi} \\
            %T11 & \url{https://zenodo.org/records/16017718} & Fangyijie et al. 2025 \\
            %T12 & \url{https://zenodo.org/records/16018021} & Jiawei et al. 2025 \\
            %T13 & \url{https://zenodo.org/uploads/16018109} & Darren Hiu Sun et al. 2025 \\
            \bottomrule
        \end{tabular}%
    }
    \label{tab:fugc2025_source_code}
\end{table}

\subsection{Top Ten Solutions}
\subsubsection{T1} Tran et al.\ adopt a human-in-the-loop semi-supervised framework for cervical ultrasound image segmentation using a U-Net architecture\cite{nam2025human}. The model is first trained on a small labeled set of 50 images using a combination of Dice and cross-entropy (CE) losses. Pseudo-labels for the unlabeled data are then generated and iteratively refined by annotators in Label Studio, with particular attention to correcting major segmentation errors such as disconnected regions and imprecise boundaries. The refined masks are progressively incorporated into the training set across multiple iterations, leading to continuous performance improvements. To ensure long-term accessibility and reproducibility, all pseudo-labels generated have been publicly released on Zenodo: \textcolor{cyan}{\href{https://zenodo.org/records/18217137}{https://zenodo.org/records/18217137}}.
\subsubsection{T2} Pham et al. introduce Fetal-BCP, a semi-supervised segmentation framework for fetal ultrasound images, addressing the distribution gap between labeled and unlabeled data using a bidirectional copy-paste strategy within a mean teacher architecture\cite{ha2025fetal}. The method employs copy-paste augmentation across labeled and unlabeled samples to generate mixed training images, followed by pseudo-label generation from the teacher model. The segmentation supervision is constructed by fusing ground truths with refined pseudo-labels through a mask-based blending mechanism. The loss function combines Dice and CE losses, with labeled and pseudo-labeled regions weighted differently. The method incorporates extensive data augmentation and post-processing, and pretraining further boosts performance.
\subsubsection{T3} Liu et al. propose a lightweight semi-supervised segmentation framework, based on UniMatch-V2, a consistency-driven model that uses pseudo-labels generated from weakly augmented inputs to supervise predictions under strong augmentations\cite{hongyu2025light}. The method enforces prediction consistency between these augmentations to enhance robustness and reduce overfitting to simple patterns. A complementary channel-wise dropout mechanism ensures diverse feature learning across dual streams. Additionally, a mean teacher model with exponential moving average (EMA) is used to stabilize pseudo-label quality. The total training loss is computed from both labeled and high-confidence unlabeled data, filtered by a threshold. The model leverages a DINOv2-Small vision transformer as the encoder and integrates strong data augmentation strategies, such as flipping, rotation, and color jittering. Designed for efficiency, the architecture is optimized via depthwise separable convolutions and attention modules.

\subsubsection{T4} Zhang et al. present a semi-supervised segmentation framework for cervical ultrasound images using a two-stage pipeline based on nnUNet and U-Net architectures\cite{chensheng2025semi}. In the first stage, nnUNet is trained on 50 gold-standard labeled images using five-fold cross-validation. This model generates initial pseudo-labels for 450 unlabeled images, which are then refined through post-processing to correct structural inconsistencies. In the second stage, U-Net models are trained on both labeled and refined pseudo-labeled data using a weighted loss function, giving higher weight to gold-standard samples. Extensive data augmentation is applied to enhance generalization. Model predictions are ensembled via a voting strategy to improve robustness. 

\subsubsection{T5} Huang et al. introduce a semi-supervised segmentation framework, leveraging UniMatch-V2 integrated with the DINOv2 self-supervised vision transformer\cite{mingxu2025leveraging}. The method addresses the limited labeled data challenge by combining consistency learning and contrastive representation learning within a unified pipeline. UniMatch-V2 replaces the dual-stream augmentation of previous versions with a single unified augmentation pipeline, improving performance and reducing computational cost. It employs complementary channel-wise dropout to enhance feature diversity by randomly suppressing and emphasizing different channels in strongly augmented views. The segmentation model is trained using high-confidence pseudo-labels generated from weakly augmented images, which are then used to supervise predictions under stronger augmentations. The DINOv2 encoder, based on a teacher-student framework updated with EMA, boosts feature representation quality. 

\subsubsection{T6} Chen et al. introduce a semi-supervised segmentation framework, which integrates co-training and Mean Teacher paradigms using heterogeneous architectures\cite{yu2025comt}. The method constructs two student-teacher pairs with distinct backbones to extract diverse features from unlabeled data. Pseudo-labels are generated using exponentially moving averaged teacher models and are reconstructed via entropy minimization, selecting the most confident predictions across models for each pixel. To further reduce noise in uncertain regions, the method introduces entropy-adaptive loss weights, applying more conservative learning to high-entropy pixels. The total loss combines supervised (Dice and CE losses) and unsupervised (weighted mean squared error-MSE) terms, balancing them with a Gaussian schedule. The framework is implemented with lightweight U-Net and DeepLabv3 as co-training networks, incorporating comprehensive augmentations. 

\subsubsection{T7} Liu et al. introduce an ensemble-based semi-supervised segmentation framework, combining PVTv2-B1 and ResNet34D within the UniMatch semi-supervised learning architecture\cite{xiao2025vision}. The method begins by training two U-Net variants using UniMatch on 50 labeled and 450 unlabeled images. High-quality pseudo-labels are then generated through inference on unlabeled data, followed by manual selection and iterative refinement in two stages, gradually expanding the labeled training set to 450 images. This enables a final round of fully supervised training using the refined pseudo-labeled dataset. The ensemble of the two backbones improves robustness by leveraging their complementary representations. A combination of Dice and CE losses is used, with data augmentation techniques such as flipping, rotation, color jitter, and CutMix applied throughout training. 

\subsubsection{T8} Xiao et al. propose a semi-supervised segmentation framework for cervical ultrasound images, combining augmentation-driven pseudo-label learning with copy-paste consistency learning within a teacher-student model\cite{yusong2025hierarchical}. Weak and strong augmentations are applied to unlabeled data, where pseudo-labels from weakly perturbed samples supervise strongly perturbed counterparts. Additionally, labeled and pseudo-labeled samples are blended via a mask-based copy-paste mechanism to form mixed training images, and consistency is enforced using corresponding mixed labels. The loss combines Dice, CE, and MSE, with distinct contributions from labeled and unlabeled regions. Extensive data augmentation, hierarchical supervision, and EMA-based teacher updates significantly improve segmentation accuracy.

\subsubsection{T9} Islam et al. introduce a transformer-based semi-supervised segmentation framework for fetal ultrasound images, leveraging the UniMatch-V2 model built upon a DINOv2 Vision Transformer encoder\cite{m2025semi}. The method employs weak-to-strong consistency learning, where the student model is trained to produce consistent predictions under weak and strong augmentations within a unified single-stream architecture. Complementary dropout at the feature level further enhances representation learning. The segmentation supervision is guided by CE and Dice losses, enabling stable training across limited labeled data and diverse unlabeled sets. The framework incorporates both internal and external unlabeled datasets and utilizes patch-wise processing with zero-padding. Pretraining with DINOv2 and multi-scale feature fusion significantly improves segmentation performance.

\subsubsection{T10} Jiang et al. introduce Semi-CervixSeg, a multi-stage semi-supervised segmentation framework, leveraging a progressive training strategy that integrates consistency regularization and contrastive learning in the initial stage\cite{juntao2025semi}. The method first applies multi-view random augmentation and enforces prediction consistency between differently augmented views to enhance generalization. A supervised loss (Dice and CE) is used on labeled data, while a mean squared consistency loss is used on unlabeled samples. Subsequently, pseudo-labels are iteratively generated and refined across multiple stages, where each stage merges labeled and pseudo-labeled samples for supervised training. The models used include RWKV-UNet and PVT-EMCAD-B2, with deep supervision applied in later stages. The training process yields increasingly accurate pseudo-labels, improving segmentation performance.

\subsection{Survey of the Methods}
The success of participants in FUGC hinges on several key aspects: preprocessing, network architecture, loss functions, data augmentation, semi-supervised learning strategies, ensemble methods, and post-processing. A detailed comparison is summarized in \textcolor[RGB]{65, 105, 210}{\textbf{Tables \ref{tab:benchmark_methods} and \ref{tab:benchmark_methods_table5}}}.

\begin{table*}[htbp]
    \centering
    \caption{Summary of the benchmark methods of top ten teams for cervical segmentation}
    \renewcommand{\arraystretch}{1.2} 
    \resizebox{\textwidth}{!}{%
        \begin{tabular}{l p{0.1\textwidth} p{0.3\textwidth} p{0.35\textwidth} p{0.85\textwidth}}
            \toprule
            \textbf{Teams} & \textbf{Framework} & \textbf{Network} & \textbf{Semi-supervised Strategy} & \textbf{Highlight Methods} \\
            \midrule
            T1 & Multi-stage & UNet & Human in the loop for Pseudo-label Generation & An annotator iterates through the prediction of unlabeled data to refine the masks using Label Studio software \\
            T2 & One-stage & DeepLabV3+ with ResNet-101 & Mean Teacher & Bidirectional consistency with Copy-Paste; filters predictions via thresholding；connected components for pseudo-label supervision\\
            T3 & One-stage & DINOv2-S & Mean Teacher Framework with Unimatch-V2 & FixMatch via multiple strong augmentation branches \\
            T4 & Two-stage & UNet, nnUNet & nnUNet for Pseudo-label Generation & Weighted loss function; Various data augmentation techniques; Postprocessing of pseudo-label \\
            T5 & One-stage & DINOv2 & Mean Teacher Framework with Unimatch-V2 & FixMatch via multiple strong augmentation branches \\
            T6 & One-stage & Lightweight UNet & Dual teacher models (UNet and DeepLabV3) & Use different model architectures to enhance diversity and apply minimum entropy to reconstruct pseudo-soft labels \\
            T7 & Three-stage & PVTv2-B1 and ResNet34D & Mean Teacher Framework with Unimatch-V2 & Integrated best models to generate pseudo labels；manually selected pseudo-labeled images to augment training data \\
            T8 & One-stage & UNet & Mean Teacher & Augmentation-driven pseudo-label learning; Copy-paste consistency learning \\
            T9 & One-stage & DINOv2-S, DINOv2-B & Mean Teacher Framework with Unimatch-V2 & FixMatch via multiple strong augmentation branches \\
            T10 & Multi-stage & RWKV-UNet & Consistency Learning and Pseudo-label Generation & Multi-view random augmentation and contrastive learning; multi-stage training to generate/optimize pseudo-labels \\
            \bottomrule
        \end{tabular}%
    }
    \label{tab:benchmark_methods}
\end{table*}

\begin{table*}[htbp]
    \centering
    \caption{Methods of top ten teams. IN: Intensity normalization; RS: Resize; EM: Ensemble model; CE: Cross-entropy.}
    \renewcommand{\arraystretch}{1.2}
    \resizebox{\textwidth}{!}{%
        \begin{tabular}{l p{0.1\textwidth} p{0.4\textwidth} p{0.85\textwidth} c c c c p{0.2\textwidth}}
            \toprule
            \textbf{Teams} & \textbf{Preprocessing} & \textbf{Postprocessing} & \textbf{Data Augmentation} & \textbf{Ensemble Model} & \textbf{Cross-Validation} & \textbf{External Data} & \textbf{Pretrained Model} & \textbf{Loss Function} \\
            \midrule
            T1 & IN & EM; Erode/dilate; Gaussian smoothing & Rotation; scaling; elastic deformations; Gamma correction; Gaussian noise; contrast/brightness adjustment; low resolution simulation & Yes & 5-fold & No & No & CE + Dice loss \\
            T2 & IN; RS & Connect components (remove outliers) & Flipping; rotation; elastic deformation; Gaussian blur/noise; brightness/contrast adjustment; CLAHE; Bidirectional Copy-Paste & No & No & No & ImageNet & CE + Dice loss \\
            T3 & IN & No & Scaling; flipping; Gaussian blur; contrast/brightness adjustment; CutMix & No & No & No & DINOv2 & CE loss \\
            T4 & IN & Connect components; hole filling; dilation; erosion & Rotation; scaling; cropping; mirror; contrast/brightness adjustment; resolution simulation; Gamma transformation & Yes & 5-fold & No & No & CE + Dice loss \\
            T5 & IN; RS & Connect components & Rotation; flipping; scaling; Gamma correction; contrast adjustment; CutMix & No & No & No & ImageNet; DINOv2 & CE + Dice loss \\
            T6 & No & Connect components; hole filling; dilation; erosion & Flipping; rotation; Gamma correction; color adjustments; Cutout (occlusion) & No & No & No & No & CE + Dice loss \\
            T7 & IN & EM & Rotation; flipping; contrast adjustment; CutMix & Yes & No & No & ImageNet & CE + Dice loss \\
            T8 & No & Connect components; hole filling & Gaussian noise; cut-out noise; DCT noise & No & 5-fold & No & No & MSE + CE + Dice loss \\
            T9 & IN & No & Flipping; Gaussian blur; CutMix; Dropout & Yes & No & PSFHS & DINOv2 & CE + Dice loss \\
            T10 & IN; RS & No & Rotation; flipping & Yes & 10-fold & No & ImageNet & MSE + CE + Dice loss \\
            \bottomrule
        \end{tabular}%
    }
    \label{tab:benchmark_methods_table5}
\end{table*}

\subsubsection{Preprocessing} Preprocessing aims to standardize inputs and reduce noise across the dataset. The majority of teams adopted intensity normalization, such as z-score or min-max scaling, to ensure uniformity in pixel value ranges. Additionally, cropping and resizing strategies were used to align image dimensions with model requirements: T5 crops to 518 × 518, T10 resizes to 384 × 384, and T2 resizes to a fixed size. In contrast, some teams, such as T1, T6, and T8, bypassed explicit preprocessing, relying on data augmentation or network robustness to handle raw input images. To further reduce background noise, several pipelines (e.g., T5 and T10) employed cervical region localization via lightweight networks before cropping the ROI.

\subsubsection{Network Architecture and Loss Function} Network architectures predominantly feature U-Net variants and vision transformers (ViT). U-Net models and their lightweight versions capitalize on skip connections to preserve spatial details, with T4 integrating nnUNet for enhanced pseudo-labeling. ViT models, such as DINOv2, effectively capture global context, with T9 indicating that DINOv2-B outperforms DINOv2-S. Hybrid architectures combine CNNs and transformers, e.g., T7 integrates PVTv2-B1 and ResNet34D, while T10 employs RWKV-UNet and PVT-EMCAD-B2. Loss functions generally combine Dice and CE, with T3 and T9 using only CE, and T8 and T10 incorporating MSE loss to ensure consistency.

\subsubsection{Data Augmentation} Data augmentation is critical for enhancing generalization across varied ultrasound data. Geometric techniques, such as rotation, flipping, and scaling, are commonly used by T1, T4, and T7, with T2 further incorporating elastic deformations. Photometric augmentations include Gaussian noise, gamma correction, and contrast adjustments, with T2 employing Contrast Limited Adaptive Histogram Equalization for handling low-contrast regions. Advanced methods extend to T2's Bidirectional Copy-Paste strategy, T5/T7/T9's CutMix, T8's Discrete Cosine Transform (DCT)-based noise suppression, and T9's Complementary Channel-Wise Dropout. One submission also experimented with elastic deformations and frequency-domain DCT-based high-frequency suppression for further enrichment.

\subsubsection{Semi-Supervised Learning Strategies}
Semi-supervised strategies form the backbone of many approaches, focusing on pseudo-labeling, consistency regularization, and human-in-the-loop refinement. Teams like T1 use manual refinement via Label Studio, iteratively improving pseudo-labels, while T4 generates and post-processes pseudo-labels with nnUNet. Consistency regularization, such as UniMatch-V2/FixMatch, is employed in models of T3, T5, and T9, guiding weak augmentations to predict strong ones. The Mean Teacher framework updates pseudo-labels through an EMA-weighted teacher-student loop, where T6 further refines labels using minimum entropy from dual architectures (i.e., U-Net and DeepLabV3). Several methods incorporate multi-stage pseudo-label refinement (e.g., T10 and T6), and human-in-the-loop strategies (i.e., T1) allow manual correction of uncertain labels to enhance accuracy.

\subsubsection{Ensemble Learning and Cross-Validation} Ensemble methods and cross-validation are integral to reducing model uncertainty and enhancing robustness. T7 combines outputs from PVTv2-B1 and ResNet34D, while T10 averages outputs from PVT-EMCAD-B2 and R50-UNet. EMA teacher-student ensembles are used in T9 for model fusion. Cross-validation is common, with T1 and T4 employing 5-fold cross-validation, and T10 using 10-fold to maximize the usage of limited labeled data. 

\subsubsection{Post-Processing} Post-processing techniques are essential for refining segmentation outputs. Common morphological operations include erosion/dilation (e.g., T1), hole filling and small region removal (e.g., T4), and Gaussian smoothing (e.g., T6). T8 retains the largest connected component to avoid isolated artifacts. Some teams, such as T1, also perform connected-component analysis to remove isolated regions and smooth boundaries.

\section{Results}
\label{sec:results}
\begin{figure}[!t]
		\centering
		\includegraphics[width=3.5 in]{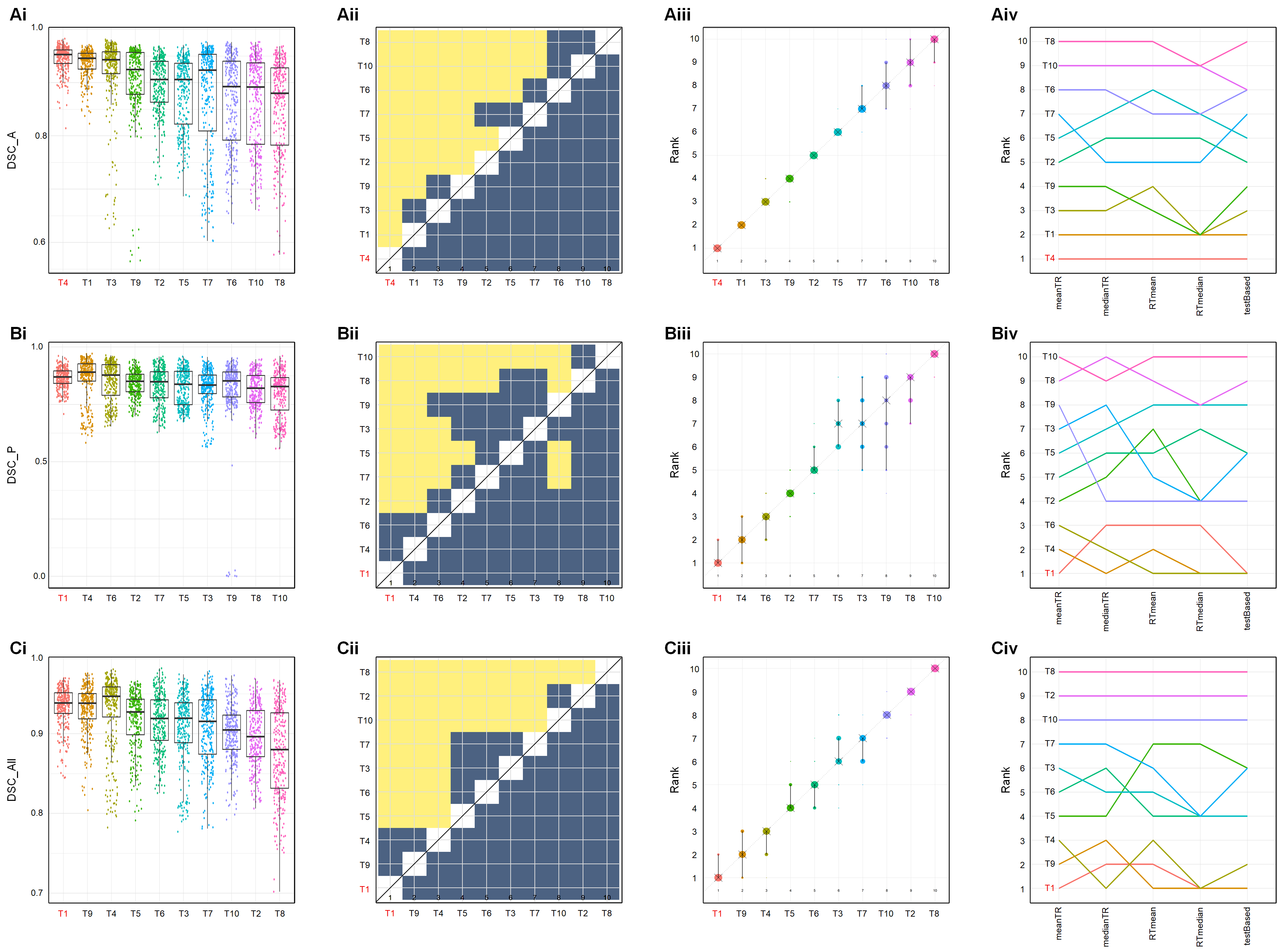}
\caption{Dice Similarity Coefficients for the anterior lip (DSC\_A), posterior lip (DSC\_P), and the overall cervix (DSC\_All). (Ai-Ci) Dot- and boxplots for visualizing the evaluation metric data. (Aii-Cii) Significance maps for visualizing the results of significance testing. (Aiii-Ciii) Blob plots for visualizing ranking stability. (Aiv-Civ) Line plots for visualizing rankings robustness across different ranking methods.}
\label{fig:fig3}
\end{figure}
\subsection{Overall Performance}
\subsubsection{DSC Metric Analysis}
\textcolor[RGB]{65,105,210}{\textbf{Fig.~\ref{fig:fig3}}} summarizes the DSC-based segmentation performance for the anterior lip (DSC\_A), posterior lip (DSC\_P), and whole lip region (DSC\_All), with numerical results provided in \textcolor[RGB]{65,105,210}{\textbf{Table~\ref{tab:dsc_hd_combined}}}. For anterior lip segmentation, T4 achieved the highest DSC\_A of 0.9457 (\textcolor[RGB]{65,105,210}{\textbf{Fig.~\ref{fig:fig3}Ai}}), and statistical significance testing (p $<$ 0.05; \textcolor[RGB]{65,105,210}{\textbf{Fig.~\ref{fig:fig3}Aii}}) confirms that T4 significantly outperformed all other teams. Ranking stability analysis further shows that T4 consistently ranked first (\textcolor[RGB]{65,105,210}{\textbf{Fig.~\ref{fig:fig3}Aiii}}), and across multiple evaluation schemes, T4 and T1 repeatedly occupied top positions, whereas other teams exhibited greater rank variability (\textcolor[RGB]{65,105,210}{\textbf{Fig.~\ref{fig:fig3}Aiv}}). For posterior lip segmentation, T1 achieved a DSC\_P of 0.8651 (\textcolor[RGB]{65,105,210}{\textbf{Fig.~\ref{fig:fig3}Bi}}), with statistical tests indicating that T1 significantly outperforms T2, T3, T5, T7, T8, T9, and T10, and performs comparably to T4 and T6 (\textcolor[RGB]{65,105,210}{\textbf{Fig.~\ref{fig:fig3}Bii}}). Ranking stability reveals that T1 fluctuates between ranks 1–2, T4 between 1–3, and T6 between 2–3 (\textcolor[RGB]{65,105,210}{\textbf{Fig.~\ref{fig:fig3}Biii}}), while evaluation schemes show T1 ranking first in testBased and RTMean, T4 leading RTMedian, MedianTR, and testBased, and T6 topping RTMean, MedianTR, and testBased (\textcolor[RGB]{65,105,210}{\textbf{Fig.~\ref{fig:fig3}Biv}}). For whole-lip segmentation, T1 again performed strongly, producing a DSC\_All of 0.9336 (\textcolor[RGB]{65,105,210}{\textbf{Fig.~\ref{fig:fig3}Ci}}), with statistical results showing that T1 significantly outperforms T2, T3, T5, T6, T7, T8, and T10, while performing comparably to T4 and T9 (\textcolor[RGB]{65,105,210}{\textbf{Fig.~\ref{fig:fig3}Cii}}). Ranking stability patterns indicate that T1 remains consistently within ranks 1–2, T9 within 1–3, and T4 within 2–3 (\textcolor[RGB]{65,105,210}{\textbf{Fig.~\ref{fig:fig3}Ciii}}), with evaluation schemes showing T1 ranking first in testBased, RTMedian, and MeanTR; T4 leading RTMedian and MedianTR; and T9 ranking first in RTMean, RTMedian, and testBased (\textcolor[RGB]{65,105,210}{\textbf{Fig.~\ref{fig:fig3}Civ}}). Overall, Teams T1, T4, T6, and T9 demonstrate superior DSC performance across anterior, posterior, and whole-lip segmentation tasks compared with the remaining teams.

\begin{table}[htbp]
    \centering
    \caption{Segmentation performance.}
    \renewcommand{\arraystretch}{1.2}
    \resizebox{\columnwidth}{!}{%
        \begin{tabular}{lcccccc}
            \toprule
            \textbf{Team} & \textbf{DSC\_All} & \textbf{DSC\_P} & \textbf{DSC\_A} & \textbf{HD\_All} & \textbf{HD\_P} & \textbf{HD\_A} \\
            \midrule
            T1  & 0.9336±0.0249 & 0.8651±0.0435 & 0.9365±0.0269 & 54.3638±31.4802 & 51.7807±28.7021 & 28.9991±26.9275 \\
            T2  & 0.8967±0.0375 & 0.8405±0.0540 & 0.8964±0.0525 & 63.9400±34.3682 & 52.9568±27.5454 & 44.3256±34.7691 \\
            T3  & 0.9080±0.0442 & 0.8215±0.0825 & 0.9191±0.0725 & 66.4719±37.2868 & 56.3524±27.8807 & 38.5524±38.6760 \\
            T4  & 0.9289±0.0458 & 0.8595±0.0964 & 0.9457±0.0239 & 52.5423±35.0126 & 48.2053±35.8816 & 29.5646±27.9845 \\
            T5  & 0.9164±0.0369 & 0.8231±0.0781 & 0.8814±0.0659 & 55.5943±32.3085 & 71.7078±49.3040 & 46.3791±32.2958 \\
            T6  & 0.9143±0.0348 & 0.8497±0.0864 & 0.8654±0.0845 & 69.8266±35.3152 & 68.3889±52.5924 & 58.8568±39.9394 \\
            T7  & 0.9070±0.0432 & 0.8317±0.0798 & 0.8718±0.1052 & 64.2765±35.6697 & 67.5738±42.0952 & 50.0431±40.8697 \\
            T8  & 0.8760±0.0560 & 0.8144±0.0733 & 0.8566±0.0878 & 80.6753±40.4531 & 61.8455±33.7392 & 63.4788±46.1018 \\
            T9  & 0.9311±0.0297 & 0.8183±0.1488 & 0.9076±0.0679 & 54.7615±27.8569 & 69.4917±42.9310 & 38.4745±26.9242 \\
            T10 & 0.9015±0.0334 & 0.7977±0.0978 & 0.8629±0.0861 & 75.6475±36.5947 & 79.7502±50.5030 & 68.2293±42.6979 \\
            \bottomrule
        \end{tabular}%
    }
    \label{tab:dsc_hd_combined}
\end{table}

\begin{figure}[!t]
		\centering
		\includegraphics[width=3.5in]{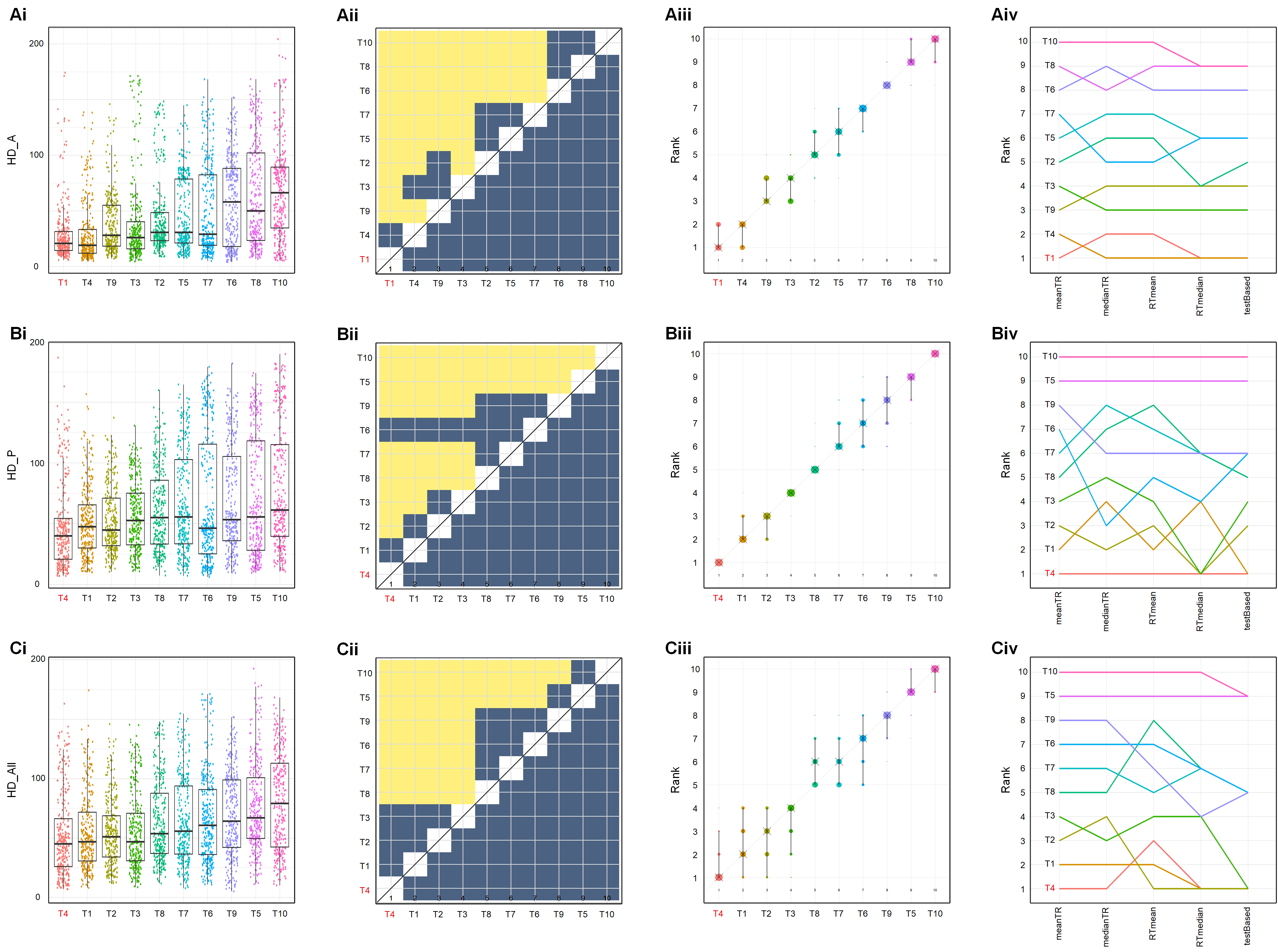}
\caption{Hausdorff Distances for the anterior lip (HD\_A), posterior lip (HD\_P), and the overall cervix (HD\_All). (Ai-Ci) Dot- and boxplots for visualizing the evaluation metric data. (Aii-Cii) Significance maps for visualizing the results of significance testing. (Aiii-Ciii) Blob plots for visualizing ranking stability. (Aiv-Civ) Line plots for visualizing rankings robustness across different ranking methods.}
\label{fig:fig4}
\end{figure}

\begin{figure}[!t]
		\centering
		\includegraphics[width=3.5 in]{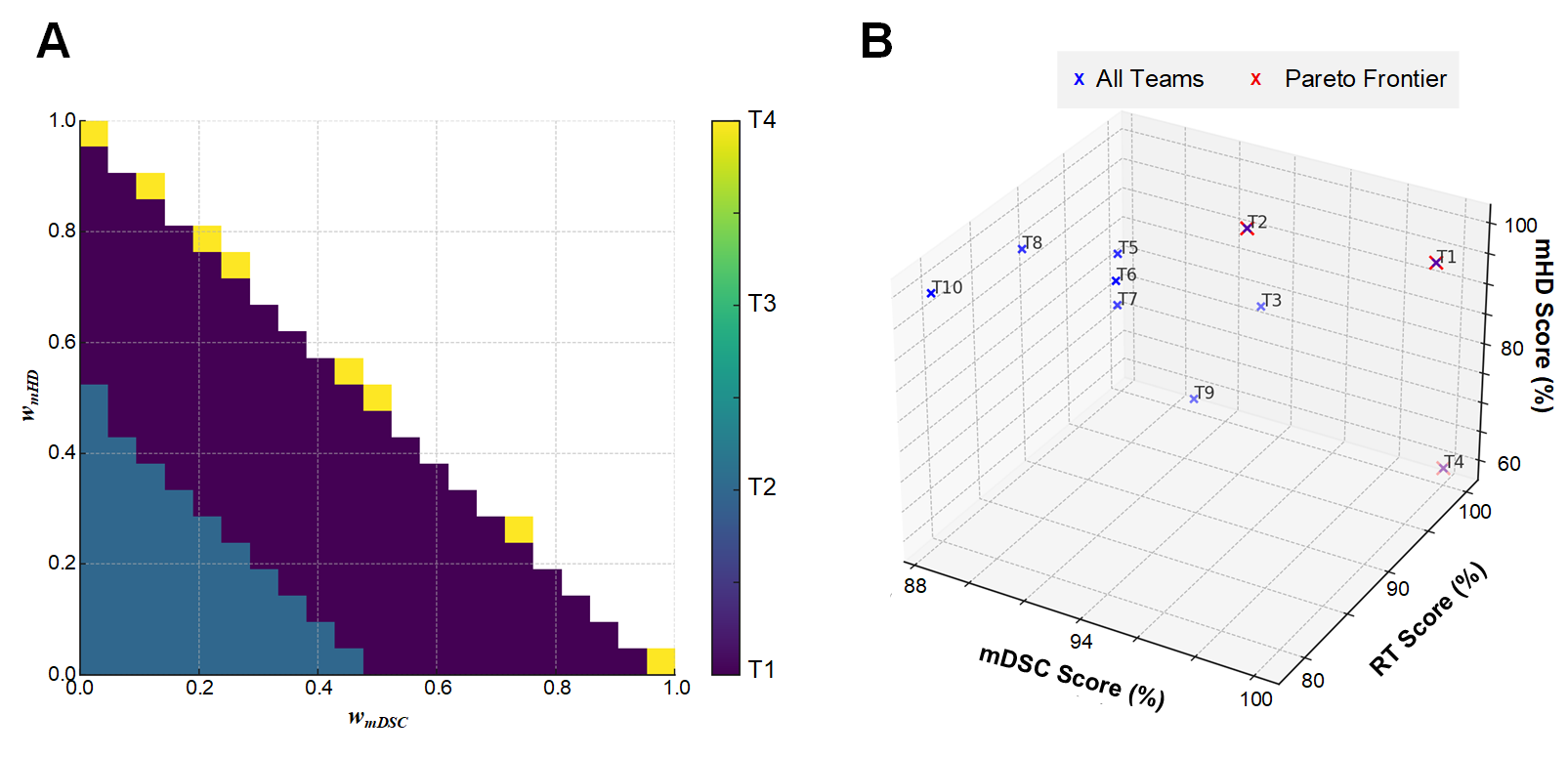}
\caption{(A) Top-model region map, showing the winning team for each $(w_{\text{mDSC}}, w_{\text{mHD}})$ pair, with $w_{\text{RT}} = 1 - w_{\text{mDSC}} - w_{\text{mHD}}$.(B) Pareto frontier illustrating trade-offs among mDSC, mHD, and RT scores.}
\label{fig:fig5}
\end{figure}

\subsubsection{HD Metric Analysis}
\textcolor[RGB]{65,105,210}{\textbf{Fig.~\ref{fig:fig4}}} summarizes the HD-based segmentation performance for the anterior lip (HD\_A), posterior lip (HD\_P), and whole-lip region (HD\_All), with corresponding numerical results reported in \textcolor[RGB]{65,105,210}{\textbf{Table~\ref{tab:dsc_hd_combined}}}. For anterior lip segmentation, T1 achieved the lowest HD\_A of 28.9991 (\textcolor[RGB]{65,105,210}{\textbf{Fig.~\ref{fig:fig4}Ai}}), and statistical testing (p $<$ 0.05; \textcolor[RGB]{65,105,210}{\textbf{Fig.~\ref{fig:fig4}Aii}}) shows that T1 significantly outperforms T2, T3, T5, T6, T7, T8, T9, and T10, while performing comparably to T4. Ranking stability indicates that both T1 and T4 remain consistently within ranks 1–2 (\textcolor[RGB]{65,105,210}{\textbf{Fig.~\ref{fig:fig4}Aiii}}), and across evaluation schemes, T1 ranks first in MeanTR, whereas T4 leads in MedianTR, RTMedian, RTMean, and testBased (\textcolor[RGB]{65,105,210}{\textbf{Fig.~\ref{fig:fig4}Aiv}}). For posterior lip segmentation, T4 attained an HD\_P of 48.2053 (\textcolor[RGB]{65,105,210}{\textbf{Fig.~\ref{fig:fig4}Bi}}), and statistical analysis reveals that T4 significantly outperforms T2, T3, T5, T7, T8, T9, and T10, while performing comparably to T1 and T6 (\textcolor[RGB]{65,105,210}{\textbf{Fig.~\ref{fig:fig4}Bii}}). T4 maintains consistently top ranking stability (\textcolor[RGB]{65,105,210}{\textbf{Fig.~\ref{fig:fig4}Biii}}) and dominates most evaluation schemes, whereas other teams exhibit substantial variability (\textcolor[RGB]{65,105,210}{\textbf{Fig.~\ref{fig:fig4}Biv}}). For whole-lip segmentation, T4 achieved the lowest HD\_All of 52.5423 (\textcolor[RGB]{65,105,210}{\textbf{Fig.~\ref{fig:fig4}Ci}}), and significance testing confirms that T4 significantly outperforms T5, T6, T7, T8, T9, and T10, while performing comparably to T1, T2, and T3 (\textcolor[RGB]{65,105,210}{\textbf{Fig.~\ref{fig:fig4}Cii}}). Ranking stability reveals that T4 generally falls within ranks 1–3, T1 and T2 within ranks 1–4, and T3 within ranks 2–4 (\textcolor[RGB]{65,105,210}{\textbf{Fig.~\ref{fig:fig4}Ciii}}). Under different evaluation schemes, T4 ranks first in MeanTR, MedianTR, RTMedian, and testBased; T1 leads in RTMedian and testBased; T2 ranks first in RTMean, RTMedian, and testBased; and T3 also secures first place in testBased (\textcolor[RGB]{65,105,210}{\textbf{Fig.~\ref{fig:fig4}Civ}}). Overall, Teams T1, T2, T3, T4, and T6 consistently demonstrate superior HD performance compared with the remaining teams.

\subsubsection{RT Metric Analysis} RT values for T1 to T10 are as follows: 315.884, 230.449, 394.435, 652.377, 107.945, 32.857, 349.566, 40.213, 509.464, and 161.377 ms. Among them, T6 and T8 achieved the lowest RTs ($<$100 ms), while T4 recorded the highest RT. T6 ranked first in terms of RT (\textcolor[RGB]{65, 105, 210}{\textbf{Table \ref{tab:team_selected_metrics}}}).     
\subsection{Ranking Results}
\textcolor[RGB]{65,105,210}{\textbf{Table~\ref{tab:team_selected_metrics}}} presents the overall ranking results evaluated using the three challenge metrics: mDSC, mHD, and RT. In terms of segmentation accuracy, T4 achieved the best performance with an mDSC of 90.2611 and an mHD of 38.8849, followed closely by T1 with an mDSC of 90.0801 and an mHD of 40.3899. Regarding runtime efficiency, T2, T5, T6, T8, and T10 received full RT scores, with T6 achieving the fastest runtime of 32.8569 ms, compared with 315.8840 ms for T1 and 652.3772 ms for T4. When considering the weighted combination of all three metrics (0.4:0.4:0.2), T1 secured first place by balancing accuracy and efficiency, whereas T4 ranked fourth due to its strong accuracy but substantially slower runtime, and T6 ranked sixth due to the opposite trade-off. To ensure that the selected weights did not bias the final leaderboard toward any particular metric or method, we conducted a comprehensive sensitivity analysis across the full weight space. A complete sweep of the weight space is illustrated in \textcolor[RGB]{65,105,210}{\textbf{Fig.~\ref{fig:fig5}A}}, where a heatmap shows the winning team for every feasible pair of $(w_{\text{mDSC}}, w_{\text{mHD}})$ with $w_{\text{RT}} = 1 - w_{\text{mDSC}} - w_{\text{mHD}}$. T1 dominates the majority of the weight space, particularly where mDSC and mHD carry moderate-to-high weights (0.3–0.5), consistent with the accuracy-focused design of this challenge, while only extreme RT-heavy settings temporarily favor the fastest models. Importantly, the chosen weighting point (0.4, 0.4, 0.2) lies at the center of a stable, accuracy-dominant region, ensuring robust rankings while preventing RT from exerting disproportionate influence. Complementing the aggregated ranking system, we further introduce a Pareto frontier analysis (\textcolor[RGB]{65,105,210}{\textbf{Fig.~\ref{fig:fig5}B}}), providing a weight-free, multi-objective assessment of the trade-offs among segmentation accuracy, boundary accuracy, and runtime. The Pareto frontier identifies T1, T2, and T4 as nondominated solutions that achieve optimal trade-offs across the three metrics.

\section{Discussion}
\label{sec:dicussion}
In the FUGC, one key finding is that the quality of the generated pseudo-labels varies significantly across different semi-supervised solutions, and competitive performance can be achieved through diverse approaches. It should be noted that specific implementation of a method holds significance and could result in different performance outcomes. For example, mean teacher framework with Unimatch-V2 were used by multiple groups\cite{hongyu2025light,mingxu2025leveraging,xiao2025vision,m2025semi}, but only T3 was ranked as a top 3. Additionally, algorithms (e.g., T1 and T4) with high segmentation accuracy often suffer from low efficiency, indicating a need for further optimization.

\begin{table*}[htbp]
    \centering
    \caption{Rankings of participating teams.}
    \renewcommand{\arraystretch}{1.2}
    \resizebox{\textwidth}{!}{%
        \begin{tabular}{p{0.2\textwidth} p{0.2\textwidth} p{0.2\textwidth} p{0.2\textwidth} p{0.2\textwidth} p{0.2\textwidth} p{0.2\textwidth}
        p{0.2\textwidth} p{0.2\textwidth}
        }
            \toprule
            \textbf{Teams} & \textbf{RT $($ms$)$}  & \textbf{mDSC$($\%$)$} & \textbf{mHD$($\%$)$} & \textbf{mDSC Score$($\%$)$} & \textbf{mHD Score$($\%$)$} & \textbf{RT Score$($\%$)$} & \textbf{Final Score$($\%$)$} & \textbf{Rank} \\
            \midrule
            T1 & 315.8840 & 90.0801& 40.3899& 99.7113 & 99.0840 & 94.9765 & 98.5134 & 1 \\
            T2 & 230.4485 & 86.8445& 48.6412& 94.5497 & 94.0618 & 100.000 & 95.4446 & 2 \\
            T3 & 394.4350 & 87.0304& 47.4523& 94.8462 & 94.7854 & 86.6897 & 93.1906 & 3 \\
            T4 & 652.3772 & 90.2611& 38.8849& 100.000 & 100.000 & 59.4777 & 91.8955 & 4 \\
            T5 & 107.9449 & 85.2269& 59.0434& 91.9692 & 87.7304 & 100.000 & 91.8798 & 5 \\
            T6 & 32.8569 & 85.7572& 63.6228& 92.8151 & 84.9430 & 100.000 & 91.1033 & 6 \\
            T7 & 349.5664 & 85.1755& 58.8084& 91.8872 & 87.8734 & 91.4231 & 90.1888 & 7 \\
            T8 & 40.2133 & 83.5487& 62.6621& 89.2920 & 85.5278 & 100.000 & 89.9279 & 8 \\
            T9 & 509.4639 & 86.2948& 53.9830& 93.6728 & 90.8104 & 74.5545 & 88.7042 & 9 \\
            T10 & 161.3766 & 83.0266& 73.9897& 88.4592 & 78.6331 & 100.000 & 86.8369 & 10 \\
            \bottomrule
        \end{tabular}%
    }
    \label{tab:team_selected_metrics}
\end{table*}
    
    \subsection{Challenge Design Choices}
In 2022, ISUOG clinical guidelines reinforced the importance of cervical length measurement using TVS for predicting spontaneous PTB, recommending three straight-line measurements between the internal and external os, with the shortest technically correct distance recorded \cite{2022ISUOG}. However, prior studies have shown that the single-line approach may underestimate the true cervical length when the cervical canal is curved, highlighting the need for anatomically reliable delineation of the anterior and posterior cervical lips \cite{xxx4056}. Accurate segmentation is therefore a prerequisite for both guideline-recommended straight-line measurements and more anatomically faithful curve-based approaches. Centering the challenge on segmentation rather than direct length estimation ensures broad clinical applicability and supports future work on more precise cervical length assessment.

Manual segmentation of cervical structures is highly labor-intensive, and labeled TVS datasets remain scarce, making semi-supervised learning particularly relevant. Yet, no established benchmark exists for semi-supervised learning in ultrasound segmentation. To address this gap, we designed the FUGC. First, participants were required to use only the official FUGC dataset, while the use of publicly available pretrained models was permitted to support stronger feature representation. Second, the submission workflow was standardized through the Codabench platform, where all evaluation metrics were computed under consistent hardware and software configurations. Third, we clearly differentiated submission policies between phases: unlimited submissions during validation enabled model refinement, whereas the test phase was restricted to seven submissions to ensure fairness. Finally, the leaderboard was frozen after the test phase, and the validation labels were released post-challenge to facilitate continued community benchmarking without affecting official rankings. Note: Participants affiliated with the organizing institutions were allowed to take part in the challenge but were not eligible for ranking-based awards.

    \subsection{Analysis of Top-Ranked Methods}
Throughout the FUGC, Teams T1, T4, and T6 consistently achieved strong results. T4 obtained the highest segmentation accuracy, T6 delivered the best efficiency, and T1 secured first place by offering accuracy close to T4 while markedly improving runtime. Methodologically, T6 adopted an online pseudo-labeling strategy without multi-stage refinement, ensembling, or cross-validation \cite{yu2025comt}, whereas T1 and T4 followed offline pseudo-labeling pipelines \cite{nam2025human,chensheng2025semi} using multi-stage retraining, cross-validation, and ensembles to enhance robustness. In terms of architecture, T1 and T4 relied primarily on U-Net variants, while T6 used a lightweight U-Net augmented with Deeplabv3 as a co-training network.

T6 further introduced a Co-Training Mean Teacher (CO-MT) framework, combining heterogeneous models with EMA-updated teachers to improve stability and pseudo-label diversity \cite{yu2025comt}. Noise suppression was achieved through entropy-based pseudo-label selection, entropy-adaptive weighting, and a hybrid loss combining CE, Dice, and weighted MSE. By contrast, T1 and T4 emphasized offline pseudo-label improvement: T1 refined pseudo-labels iteratively using Label Studio with 5-fold cross-validation to progressively enhance label quality \cite{nam2025human}, whereas T4 employed a two-stage pipeline—nnUNet for initial pseudo-label generation and automated post-processing, followed by training U-Net models with loss reweighting to account for label reliability \cite{chensheng2025semi}. Both teams used model ensembling to obtain stable final predictions.

In summary, T1 and T4 share common principles—offline pseudo-labeling, staged training, cross-validation, and ensembling—but differ in refinement depth, label-quality handling, and architectural design. T1 depends on human-in-the-loop correction, while T4 applies automated refinement with explicit label weighting. T6 demonstrates the strength of an efficient online framework. Together, these strategies highlight the value of integrating high-quality annotations with well-designed pseudo-labeling to advance semi-supervised cervical segmentation.

\subsection{Algorithmic Design Choices}
Although direct ablation studies were not feasible due to the methodological diversity across submissions, several clear performance trends emerged. Among the top teams, offline pseudo-labeling—adopted by T1 and T4—consistently produced the highest segmentation accuracy and the lowest boundary error. Both teams enhanced pseudo-label quality through systematic post-processing, with T1 refining boundaries using smoothing and morphological operations, while T4 emphasized anatomical plausibility by filling incomplete regions and removing small disconnected components. In contrast, online approaches, despite considerable variation in augmentation and architectural design, shared a common reliance on teacher–student consistency. These ranged from UniMatch-V2 with cross-augmentation consistency to enhanced Mean Teacher variants, and T6 further integrated co-training with entropy-guided pseudo-label reconstruction. Collectively, these results demonstrate that well-designed augmentation strategies, explicit modeling of label uncertainty, and architectural diversity can produce strong semi-supervised performance, although offline refinement remains the most robust pathway for maximizing segmentation accuracy.

Across the top-performing teams, several shared design elements also contributed to performance gains. Most teams employed z-score normalization, extensive spatial and intensity augmentations, and post-processing procedures such as smoothing, artifact removal, and region completion to improve prediction reliability. Composite loss functions were widely used to balance pixel-level and structural optimization, and pretrained models were commonly leveraged to enhance feature representation. In our additional experiments (\textcolor{cyan}{\href{https://github.com/baijieyun/ISBI-2025-FUGC-Source-Code}{https://github.com/baijieyun/ISBI-2025-FUGC-Source-Code}}), DINOv3 consistently outperformed DINOv2, achieving a higher overall Dice score (DSC\_All = 0.9103 vs.\ 0.8731) and a markedly better posterior-region accuracy (DSC\_P = 0.7979 vs.\ 0.6990), together with lower overall and posterior Hausdorff distances (HD\_All = 66.48 vs.\ 73.67, HD\_P = 81.61 vs.\ 89.56), indicating improved boundary localization in challenging regions. These findings demonstrate the benefit of more advanced foundation models. Ensemble strategies and cross-validation were frequently applied to improve robustness and reduce model variance. These observations align with ablation evidence from participating teams: T1 showed that augmentation improved DSC by 1.52\% and reduced HD by 26 pixels \cite{nam2025human}; T2 demonstrated substantial gains from pretraining \cite{ha2025fetal}; and T9 confirmed that feature-level augmentations in UniMatch-V2 provided more consistent performance on challenging substructures \cite{m2025semi}. Together, these findings highlight that beyond the choice of semi-supervised framework, the careful engineering of preprocessing, augmentation, loss formulation, and model initialization plays a crucial role in achieving optimal segmentation performance.

Segmentation efficiency was also evaluated to reflect the requirements of real clinical deployment. Teams explored a wide spectrum of architectures to balance accuracy and computational cost. Lightweight U-Net variants—used by T1, T4, T6, and T8—offered strong accuracy–efficiency trade-offs, with T6 and T8 achieving the fastest inference times among all submissions. Meanwhile, other teams pursued more complex or pretrained backbones, such as DeepLabV3+ (T2) and RWKV-UNet (T10), which blend convolutional and attention-based mechanisms. Despite their architectural differences, the most efficient methods consistently relied on streamlined U-Net designs, underscoring that compact U-Net–style architectures remain highly competitive for real-time or near-real-time clinical applications. These findings demonstrate that both architectural simplicity and thoughtful model optimization remain central to achieving robust, deployable segmentation performance.

\subsection{Impact and Clinical Adoption}
This challenge addresses the ultrasound-based anatomical structure segmentation tasks essential for PTB prediction, as outlined in the ISUOG guidelines \cite{2022ISUOG}, while considering practical clinical requirements such as high accuracy and near--real-time performance. To our knowledge, it represents the first semi-supervised image analysis challenge dedicated specifically to ultrasound, and its outcomes provide important insights for advancing semi-supervised methodology in this domain.

Across the top-performing solutions, several best practices clearly emerged. Offline pseudo-labeling with multi-stage refinement proved highly effective, whereas teacher--student consistency formed the backbone of strong online approaches. High-performing teams also shared common engineering practices, including z-score normalization, diverse augmentation strategies, and composite CE + Dice loss functions to jointly optimize pixel-level fidelity and structural coherence. Pretrained feature extractors such as ImageNet-based encoders and DINOv2 substantially enhanced representation quality, while post-processing procedures (e.g., smoothing, artifact removal, and region completion) further improved anatomical plausibility. Ensemble inference and cross-validation contributed additional robustness. To satisfy clinical runtime constraints, many teams adopted lightweight U-Net variants that maintained competitive accuracy while achieving the fastest inference speeds. Taken together, these findings indicate that integrating high-quality pseudo-label refinement, consistency regularization, strong augmentation, well-designed loss functions, pretrained features, and computationally efficient architectures provides a practical and effective recipe for semi-supervised cervical segmentation.

The adoption of these learning-based strategies reflects a promising direction for future development. Many longstanding barriers in semi-supervised ultrasound segmentation have been significantly reduced, suggesting strong potential for clinical translation. Notably, the top-ranking methods achieved DSC\_All scores above 93\%, approaching the 90--95\% performance range of state-of-the-art fully supervised algorithms \cite{wlodarczyk2019estimation,wlodarczyk2020preterm,dagle2022automated,pegios2023preterm,bones2024automatic,dagle2025generating,kwon2025deep}. Offline pseudo-labeling also enables meaningful clinician--algorithm interaction, allowing experts to validate and refine uncertain predictions when necessary. In terms of latency, leading approaches achieved inference times as low as 32~ms, moving increasingly toward real-time applicability\cite{ou2024rtseg}. Finally, although reliable pseudo-label generation remains a key challenge in deep learning for medical imaging, the results of this challenge demonstrate that well-designed semi-supervised pipelines can substantially mitigate this limitation. Continued work on label quality, uncertainty modeling, and efficient deployment will further enhance the clinical readiness of these algorithms.

\subsection{Limitations and Future Directions}
This study has several important limitations that also motivate future work. First, all reference annotations were provided by a single ultrasound expert with over 10 years of experience, which ensured high-quality labeling but limited the assessment of inter-observer variability and annotation diversity; future studies will involve multiple annotators to improve robustness and generalizability. Second, although the top-performing methods achieved high segmentation accuracy, their processing times typically exceeded 200 ms per frame, and our real-time evaluation based on a 402 ms threshold on an NVIDIA Jetson Nano platform explicitly reflects the trade-off between model accuracy and real-time deployability. While the current results demonstrate practical feasibility in constrained clinical hardware, further optimization will be required for strict real-time deployment. Finally, future extensions of this benchmark will include more diverse clinical cohorts, particularly preterm cases, together with multi-rater annotations and systematic evaluation on edge computing platforms to further improve clinical applicability.

	\section{Conclusion}
	\label{sec:conclusion}
This paper presents a comprehensive review of the methods submitted to the FUGC, the first benchmark to evaluate semi-supervised cervical segmentation in transvaginal ultrasound imaging. Despite having only 50 labeled samples, all participating teams effectively leveraged the combination of limited annotations and abundant unlabeled data, with top-performing solutions adopting strategies such as pseudo-label refinement, multi-stage training, and ensemble learning. To support transparency and accelerate progress in this field, all datasets, annotations, and source code from the challenge have been publicly released, providing a reproducible platform for further research. While encouraging, the study is limited by single-expert annotations and the lack of validation on edge devices, which should be addressed in future work. 
\label{sec:references}
\bibliographystyle{IEEEtran}
\bibliography{myref}
\end{document}